# PSR J0045–7319: A DUAL-LINE BINARY RADIO PULSAR

J. F. Bell, M. S. Bessell, B. W. Stappers

*Mount Stromlo and Siding Spring Observatories, Institute of Advanced Studies,*
*Australian National University, Private Bag, Weston Creek, ACT 2611, Australia*
*email: bell@mso.anu.edu.au, bessell@mso.anu.edu.au, bws@mso.anu.edu.au*

M. Bailes

*Australia Telescope National Facility, CSIRO, PO Box 76, Epping, NSW 2121, Australia*
*email: mbailes@atnf.csiro.au*

V. M. Kaspi[1]

*IPAC/Caltech/Jet Propulsion Laboratory, MS 100-22, Pasadena, CA 91125*
*email:vicky@ipac.caltech.edu*

## ABSTRACT

Binary radio pulsars are superb tools for mapping binary orbits, because of the precision of the pulse timing method (e.g Taylor & Weisberg 1989). To date, all orbital parameters for binary pulsars have been derived from observations of the pulsar alone. We present the first observations of the radial velocity variations due to the binary motion of a companion to a radio pulsar. Our results demonstrate that the companion to the Small Magellanic Cloud pulsar PSR J0045–7319 is the B1V star identified by Kaspi et al. (1994). The mass ratio of the system is $6.3 \pm 1.2$, which, for a neutron star mass of 1.4 $M_\odot$, implies a mass of $8.8 \pm 1.8$ $M_\odot$ for the companion, consistent with the mass expected for a B1V star. The inclination angle for the binary system is therefore $44° \pm 5°$, and the projected rotational velocity of the companion is $113 \pm 10$ km s$^{-1}$. The heliocentric radial velocity of the binary system is consistent with that of other stars and gas in the same region of the Small Magellanic Cloud.

*Subject headings:* Pulsars:individual PSR J0045–7319, Stars: fundamental parameters, Stars: binaries: spectroscopic

---

[1] Hubble Fellow

The only known pulsar in the Small Magellanic Cloud, PSR J0045−7319, was discovered in a systematic search for radio pulsars in the Magellanic Clouds (McConnell et al. 1991). It was recently shown to be in a 51-day binary orbit of eccentricity 0.8 (Kaspi et al. 1994). The pulsar's mass function of 2.17 $M_\odot$ implies a minimum mass for the companion of 4.0 $M_\odot$ for a 1.4-$M_\odot$ pulsar. Since the maximum mass of a neutron star is ∼ 3.0 $M_\odot$ (Friedman, Ipser & Parker 1984, Friedman et al. 1988), the companion must be either a black hole or a massive non-degenerate star. The B1V star identified at the position of the pulsar is a strong candidate for the companion (Kaspi et al. 1994).

For a B1V star companion, several observable effects are expected, since the pulsar approaches to within approximately 5 stellar radii from the star at periastron. At radio wavelengths, effects that would vary with orbital phase include dispersion, scattering and absorption of the pulsed emission. Not one of these effects has been observed, however systematic frequency-independent timing residuals with respect to a Keplerian orbit have been detected (Kaspi et al. 1995). Optically, the star should show radial velocity variations with an amplitude of approximately 30 km s$^{-1}$. It has also been suggested that the companion may exhibit flux variations due to stellar pulsations excited resonantly by the eccentric orbit (Kumar, Ao & Quataert 1995). While the 0.01 magnitude amplitude of this effect will be very difficult to detect, such small variations have been detected in other objects (Balona 1995). In contrast, if the true companion is a black hole and the B1V star is not associated, none of the above effects should be observed (Lipunov, Postnov & Prokhorov 1995).

To determine the nature of the companion to PSR J0045−7319 observations of the B1V star were conducted to search for radial velocity variations. Spectra at a resolution of 0.55Å per pixel, covering 500Å centered on 3900Å, were obtained at nine epochs using the Australian National University's 2.3-m telescope at Siding Spring, NSW, Australia. The [OII] doublet at 3728Å from the surrounding HII region was used as a radial velocity standard. Helium-argon arcs were used to determine the dispersion. Radial velocities were obtained by Fourier cross correlating spectra from each epoch with a spectrum obtained by summing the spectra from all epochs. The spectral region used contained the Balmer lines between the Balmer jump and H$\delta$.

With the binary period, longitude and epoch of periastron, and eccentricity determined from radio observations (Kaspi et al. 1995), the systemic radial velocity and amplitude of the companion's radial velocity variation were determined using a least-squares fit to the observed velocities. The fit to the radial velocity curve is shown in Figure 1 along with the residuals and the pulsar's radial velocity curve. This fit gives an apparent semi-major axis for the companions orbit of 27.7 ± 5.0 lt-sec and hence, a mass ratio of 6.3 ± 1.2. Since all measurements of the mass of neutron stars are consistent with 1.4 $M_\odot$ (Thorsett et al. 1993), it assumed that the mass of the neutron star in this system is also 1.4 $M_\odot$. Using the pulsar mass function of $f = 2.17 M_\odot$, implies a companion mass of 8.8 ± 1.8 $M_\odot$ and an inclination angle for the binary of 44° ± 5°. The derived mass is close to the typical mass of a B1V star, ∼11 $M_\odot$ (Andersen 1991). Using the Stefan-Boltzmann law, with an effective temperature of 24000 ± 1000 K and luminosity of $1.2 \times 10^4$ $L_\odot$, the radius of the companion is 6.4 ± 0.7 $R_\odot$.

To test the stability of the fit, the fitting procedure was repeated with the point just after orbital phase zero removed, resulting in an increase of the mass ratio of only 12%. The probability that the results shown here are due to random radial velocity variations in an isolated B star is very small, considering the close positional coincidence (Kaspi et al. 1994), the good agreement of the orbital phases of the radio and optical radial velocity curves, and the amplitude of the optical radial velocity curve which gives a plausible estimate for the mass of the B star. If we assume the points are randomly distributed about a mean value, the reduced chi-squared of the prefit data is 4.4, while that of the postfit data is 0.9. For 8 degrees of freedom, the probability of obtaining a chi-squared as high or higher than 4.4 is less than 0.001, while the probability of obtaining a chi-squared as high or higher than 0.9 is 0.5. The accuracy of the measurements here could be improved with better resolution and higher signal to noise ratio.

A high rotational velocity for the companion might be expected as evolutionary histories for this type of system involve mass transfer from the pulsar progenitor to the companion, which serve to spin it up (Bhattacharya & van den Heuvel 1991). Using the He I lines, the projected rotational velocity for the companion is 113 ± 10 km s$^{-1}$. Such velocities however are common amongst main-sequence B1 stars. As-suming that the spin and orbital angular momenta are aligned the rotational angular velocity of the companion at its equator is $v/r = 3.1 \pm 0.3$ radians per day. When compared with the orbital angular velocity at periastron of $2\pi P_b^{-1}(1+e)^{0.5}(1-e)^{-1.5} = 1.96$ radians per day, this indicates that synchronisation is unlikely to have occured, although it cannot be ruled out given the uncertainty in the B star's inclination relative to the orbit.

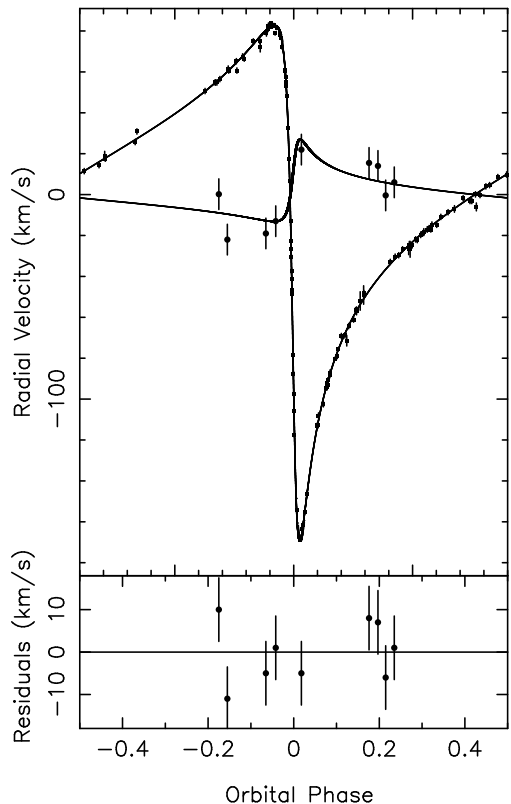

Fig. 1.— Top: Radial velocity data and fitted curves for the radio observations (Kaspi et al. 1995) (large amplitude curve) and for the companion, from optical observations (small amplitude curve). Bottom: Residuals of the two-parameter fit to the optical radial velocity variations of the companion. In both panels the error bars shown are ± one standard deviation.

The radial velocity of the companion was found to be $-10 \pm 7$ km s$^{-1}$ relative to the surrounding nebula NGC 248 (N13A + N13B) (Henize 1956). The heliocentric radial velocity of $127 \pm 7$ km s$^{-1}$ for NGC 248 compares well with the previously measured value of $131 \pm 3$ km s$^{-1}$ (Smith & Weedman 1973). The binary system's observed heliocentric velocity of $117 \pm 7$ km s$^{-1}$ is consistent within the uncertainties of the $167 \pm 40$ km s$^{-1}$ estimate from low dispersion spectra (Kaspi et al. 1994). The observed heliocentric velocities of both the PSR J0045−7319 binary and NGC 248 are somewhat less than the mean heliocentric radial velocity for the Small Magellanic Cloud (SMC) of 160 km s$^{-1}$ (Martin, Maurice & Lequeux 1989). However, when one considers the radial velocities of the four components of the SMC, −45, −28, 9 and 30 km s$^{-1}$ (relative to the mean) (Martin, Maurice & Lequeux 1989), the difference is not surprising. Observed heliocentric radial velocities for M stars (Maurice et al. 1987) in the same region as PSR J0045−7319 are similar to that of the pulsar.

The results outlined here demonstrate that the B1V star identified by Kaspi et al. (1994) is the companion to PSR J0045−7319, and rule out a black-hole companion proposed by Lipunov et al. (1995). This makes PSR J0045−7319 the first dual-line binary radio pulsar. The fact that no known radio pulsar has a black-hole companion means that we have no definitive proof that such systems exist, despite the existence of objects such as Cygnus X-3 which might be expected to form a binary system comprised of a neutron star and a black-hole in the future (van Kerkwijk et al. 1992). This might be because binaries containing black holes are particularly susceptible to coalescence during the final common-envelope phase of their evolution. However, the expected number of observable pulsar black-hole binaries, given the statistics of the known pulsar population, is only of the order of one (Narayan, Piran & Shemi 1991, Lipunov et al. 1994). A further implication of the companionship of the pulsar and B star is that the wind of the companion is tenuous compared with those of typical Galactic B stars, as was suggested by Kaspi et al. (1994). Selection may play a role here, as higher stellar-wind densities might have rendered the pulsar invisible as in the case of PSR B1259−63 near periastron (Johnston et al. 1992).

We thank Ken Freeman, Dick Manchester, Alex Rodgers, Kim Sebo and Marten van Kerkwijk for helpful discussions. We also thank the referee Fred Rasio for helpful suggestions. JFB received support from an APRA and the ATNF student program. BWS received support from an ANU scholarship. MB is a QEII Fellow. VMK received support from NASA via a Hubble Fellowship through grant number HF-1061.01-94A from the STScI, which is operated by the Association of Universities for Research in Astronomy, Inc., under NASA contract NAS5-26555. Part of this research was carried out at the JPL, California Institute of Technology, under contract with the NASA.## REFERENCES

Andersen, J. 1991, Astr. Astrophys. Rev., 3, 91.

Balona, L. A. 1995, in New Developments in Array technology and Applications, IAU Symp. 167, ed. A. G. D. Philip, K. A. Janes, and A. R. Upgren, Kluwer Dordrecht, in the press.

Bhattacharya, D. and van den Heuvel, E. P. J. 1991, Phys. Rep., 203, 1.

Friedman, J. L., Ipser, J. R., Durisen, R. H., and Parker, L. 1988, Nature, 336, 560.

Friedman, J. L., Ipser, J. R., and Parker, L. 1984, Nature, 312, 255.

Henize, K. G. 1956, ApJS, 2, 315.

Johnston, S., Manchester, R. N., Lyne, A. G., Bailes, M., Kaspi, V. M., Qiao, G., and D'Amico, N. 1992, ApJ, 387, L37.

Kaspi, V. M., Johnston, S., Bell, J. F., Manchester, R. N., Bailes, M., Bessell, M., Lyne, A. G., and D'Amico, N. 1994, ApJ, 423, L43.

Kaspi, V. M., Manchester, R. N., Bailes, M., and Bell, J. F. 1995, in Compact Stars in Binaries, IAU Symp. 165, ed. J. van Paradijs, E.P.J. van den Heuvel, and E. Kuulkers, Kluwer Dordrecht, in the press.

Kumar, P., Ao, C. O., and Quataert, E. J. 1995, Astrophys. J. in the press.

Lipunov, V. M., Postnov, K. A., and Prokhorov, M. E. 1995, ApJ, 441, 776.

Lipunov, V. M., Postnov, K. A., Prokhorov, M. E., and Osminkin, E. Y. 1994, ApJ, 423, L121.

Martin, N., Maurice, E., and Lequeux, J. 1989, A&A, 215, 219.

Maurice, E. et al. 1987, Astr. Astrophys. Suppl. Ser., 67, 423.

McConnell, D., McCulloch, P. M., Hamilton, P. A., Ables, J. G., Hall, P. J., Jacka, C. E., and Hunt, A. J. 1991, MNRAS, 249, 654.

Narayan, R., Piran, T., and Shemi, A. 1991, ApJ, 379, L17.

Smith, M. G. and Weedman, D. W. 1973, ApJ, 179, 461.

Taylor, J. H. and Weisberg, J. M. 1989, ApJ, 345, 434.

Thorsett, S. E., Arzoumanian, Z., McKinnon, M. M., and Taylor, J. H. 1993, ApJ, 405, L29.

van Kerkwijk, M. H. et al. 1992, Nature, 355, 703.
This 2-column preprint was prepared with the AAS LaTeX macros v3.0.4